\documentclass[twocolumn,showpacs,aps,superscriptaddress,%
  eqsecnum,prd,notitlepage,showkeys,10pt]{revtex4-1}

\usepackage{amssymb}
\usepackage{amsmath}
\usepackage{graphicx}
\usepackage{dcolumn}
\usepackage{hyperref}
\usepackage{upgreek}

\begin{document}

\title{Selected topics from the single top \boldmath$t$-channel:\\ cross section and other properties}

\author{Benedikt Maier}
\affiliation{Karlsruhe Institute of Technology (KIT) -- D-76131 Karlsruhe, Germany \\ \vspace{-0.3cm}${}$\\on behalf of the ATLAS and CMS collaborations.\\\vspace{0.2cm} ${}$\\Preprint of the proceedings for the contribution to the LHCP2015 conference, St.\,Petersburg, Russia.}

\begin{abstract}
Measurements of the cross section and of the interactions happening at the tWb vertext are performed in the single top $t$-channel at center-of-mass energies of $\sqrt{s}=7$ and $8$\,TeV. Results of both ATLAS and CMS collaborations are presented. No indications for new physics and no deviations from the Standard Model predictions within the experimental and theoretical uncertainties are found.
\end{abstract}

\maketitle

\section{INTRODUCTION}

\begin{figure}
\includegraphics[width=0.2\textwidth]{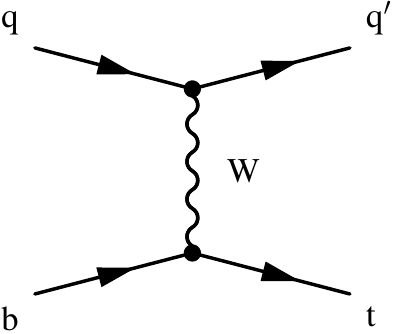}\hspace{0.4cm}
\includegraphics[width=0.2\textwidth]{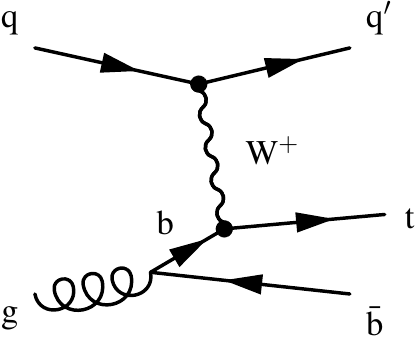}
\caption{Leading order Feynman diagram of the $t$-channel in the 5F (left) and the 4F (right).}
 \label{fig:feyn}
\end{figure}

The single top $t$-channel has the largest cross section of the three modes the electroweak production of a  top quark at hadron colliders typically is devided into. The separation into $t$-, tW- and $s$-channel makes most sense at leading order in the strong coupling constant $\alpha_\mathrm{S}$. At next-to-next-to-leading order at the latest, the definitions are not unambiguous anymore, and $t$- and $s$-channels start to interfere. It therefore makes it an interesting place to look for potential new structures in the tWb coupling and to measure key parameters of the Standard Model (SM) such as the CKM matrix element $V_\mathrm{tb}$ that, in contrast to top quark pair production which is mediated by the strong interaction, appears already in the production. The leading order Feynman diagram of Figure~\ref{fig:feyn} moreover suggests that the rates for the production of a top quark (t) are different from the anti-top ($\bar{\mathrm{t}}$) quark in proton-proton collisions, because the incoming light quark is more likely to be a quark than its anti-partner. In turn this means the $t$-channel can also be used to constrain parton distribution functions (PDFs), which each predict a different $R=\sigma_\mathrm{t}/\sigma_{\bar{\mathrm{t}}}$ according to the respective energy and momenta  distributions the partons are carrying.

\section{CHANNEL TOPOLOGY AND ANALYSIS STRATEGY} 

Single top $t$-channel events have a distinct topology explained in the following: most characteristic is the upper outgoing quark line in Figure~\ref{fig:feyn}, representing a light quark which recoils against the exchanged virtual W boson. Upon hadronization it results in light jet with substantial transverse momentum, which tends to go in a forward direction. In a typical analysis the resonant top quark is required to decay leptonically, rejecting multi-jet background processes for which it is difficult to fake a prompt lepton. The sign of the lepton will also be used for distinguishing between t and $\bar{\mathrm{t}}$ production. The top decay also features a b~quark giving rise to a central b~jet. 
The initial b~quark is implied to stem from a gluon splitting. The $\bar{\mathrm{b}}$ quark and its corresponding b jet however lie out of the tracker acceptance most of the time and thus cannot be tagged. Depending on whether one chooses the 5- or 4-flavor-scheme (5F, 4F) to describe the proton (in the latter the b quark is not considered a massless parton, but must be produced in a gluon splitting), the leading order formulation in terms of Feynman diagrams and calculus is either a $2\to2$ or $2\to3$ process. This has deep implications for the predictions and the modelling of the $t$-channel: for an all-orders-expansion the two schemes must give exactly the same results; in practice simulations at next-to-leading order are employed, and this circumstance leads to different predictions for the 4F and 5F. These can then be compared to experimental data, and conclusions can tried to be drawn on which model is the better. A detailed theoretical introduction into this subject is provided in~\cite{maltoni}.

\newcommand{\pt}{\ensuremath{p_\mathrm{T}}}
\newcommand{\ttbar}{\ensuremath{\mathrm{t}\bar{\mathrm{t}}}}
\newcommand{\ttbarH}{\ensuremath{\mathrm{t}\bar{\mathrm{t}}\mathrm{H}}}
\newcommand{\abseta}{\ensuremath{\mid\hspace{-0.085cm}\eta\hspace{-0.085cm}\mid\,\,}}

The topology described above lends itself to a so called ``2 jets 1 tag" selection (2j1t), which is widely employed in single top measurements and also consistently applied in every analysis presented here. Besides an isolated, hard lepton one expects one jet identified as b~jet and a light, forward jet. In this signal enriched phase space it is typically the disciminator of a Neural Network trained with variables separating between single top production and the main background processes (\ttbar, W + jets and multi-jet production) or the pseudorapidity of the untagged jet, $\mid\hspace{-0.085cm}\eta_\mathrm{j'}\hspace{-0.085cm}\mid$, that is used to extract the signal. The correct modelling of backgrounds is often verified in 2j0t or 3j2t control regions which are enriched in W + jets and \ttbar~events, respectively. This guarantees that all analysis ingredients are validated in phase spaces which are very close yet entirely orthogonal to the signal region.

\section{CROSS SECTIONS}
\subsection{Inclusive}

The CMS cross section measurement at $\sqrt{s}=8$\,TeV~\cite{cms8tev} is designed as a template analysis in $\mid\hspace{-0.085cm}\eta_\mathrm{j'}\hspace{-0.085cm}\mid$, the pseudorapidity of the light recoil jet. It selects events with exactly one muon (electron) with $\pt>26$\,GeV (30\,GeV) and $\abseta<2.1$ ($2.5$). Events with additional lepton candidates with looser selection criteria are rejected. For the muon channel a cut on the transverse mass of the reconstructed W boson $m_\mathrm{T}>50$\,GeV is imposed. The definition for the transverse mass is $m_\mathrm{T}=\sqrt{(p_\mathrm{T}^{\upmu}+E_\mathrm{T}^\mathrm{miss})^2-(p_x^\upmu+p_x^\mathrm{miss})^2-(p_y^\upmu+p_y^\mathrm{miss})^2}$. It relies on the missing transverse energy components to balance the sum of all observed momenta. In the electron channel it is a cut on the missing transverse energey of $E_\mathrm{T}^\mathrm{miss}>45$\,GeV that helps reject the QCD multi-jet background. By means of a range for the reconstructed top quark mass, which is $130<m_{\ell\upnu\mathrm{b}}<220$\,GeV, a signal (inside) and sideband (outside) region is defined.

The analysis exploits a 3j2t control region to determine the \ttbar~contribution in a semi-data driven way by looking at the $\abseta$templates of the untagged jet. Contributions for all other SM processes except for \ttbar~are subtracted from the data template, and bin-by-bin correction factors are derived by dividing the observed yields by the \ttbar~prediction as taken from simulation. This set of correction factors is then applied to the \ttbar~template in the 2j1t region, both in the signal and sideband regions. Since events in the 2j0t sample are predominantly stemming by W + jets production and hence these events have jets mostly coming from light quarks, this region is only used to perform a general validation of W + jets shapes and it is instead preferred to derive bin-by-bin correction factors for this background from the sidebands in 2j1t region. Predicted yields from all other processes are subtracted from the data $\mid\hspace{-0.085cm}\eta_\mathrm{j'}\hspace{-0.085cm}\mid$ template and scale factors with respect to the W + jets simulation are obtained. The simulation is also used to derive additional corrections by extrapolating from the sideband into the signal region. QCD contributions are derived in a purely data driven manner from a region with inverted criteria on lepton isolation, but turn out to be very small.

\begin{figure}
\includegraphics[width=0.38\textwidth]{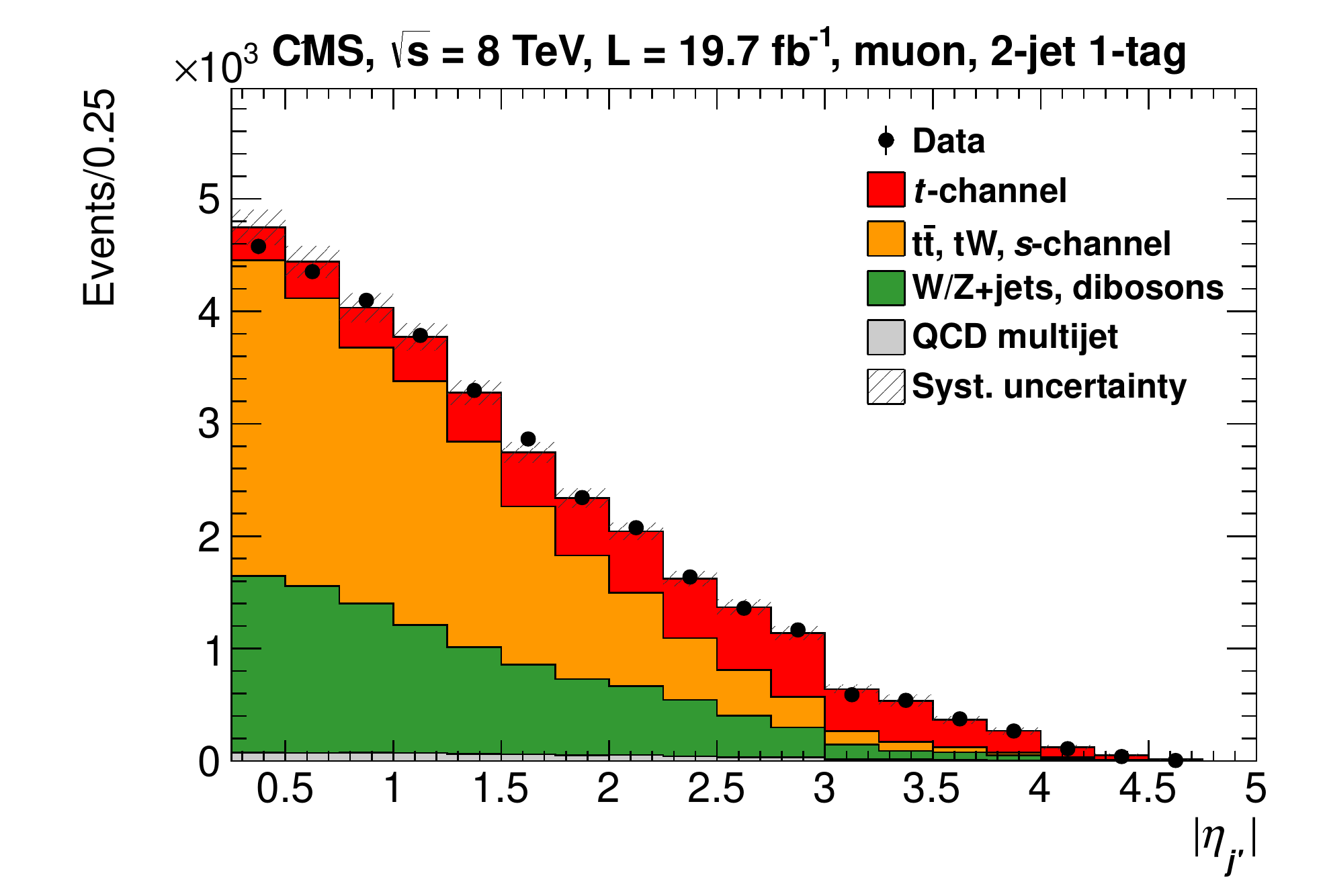}
\includegraphics[width=0.38\textwidth]{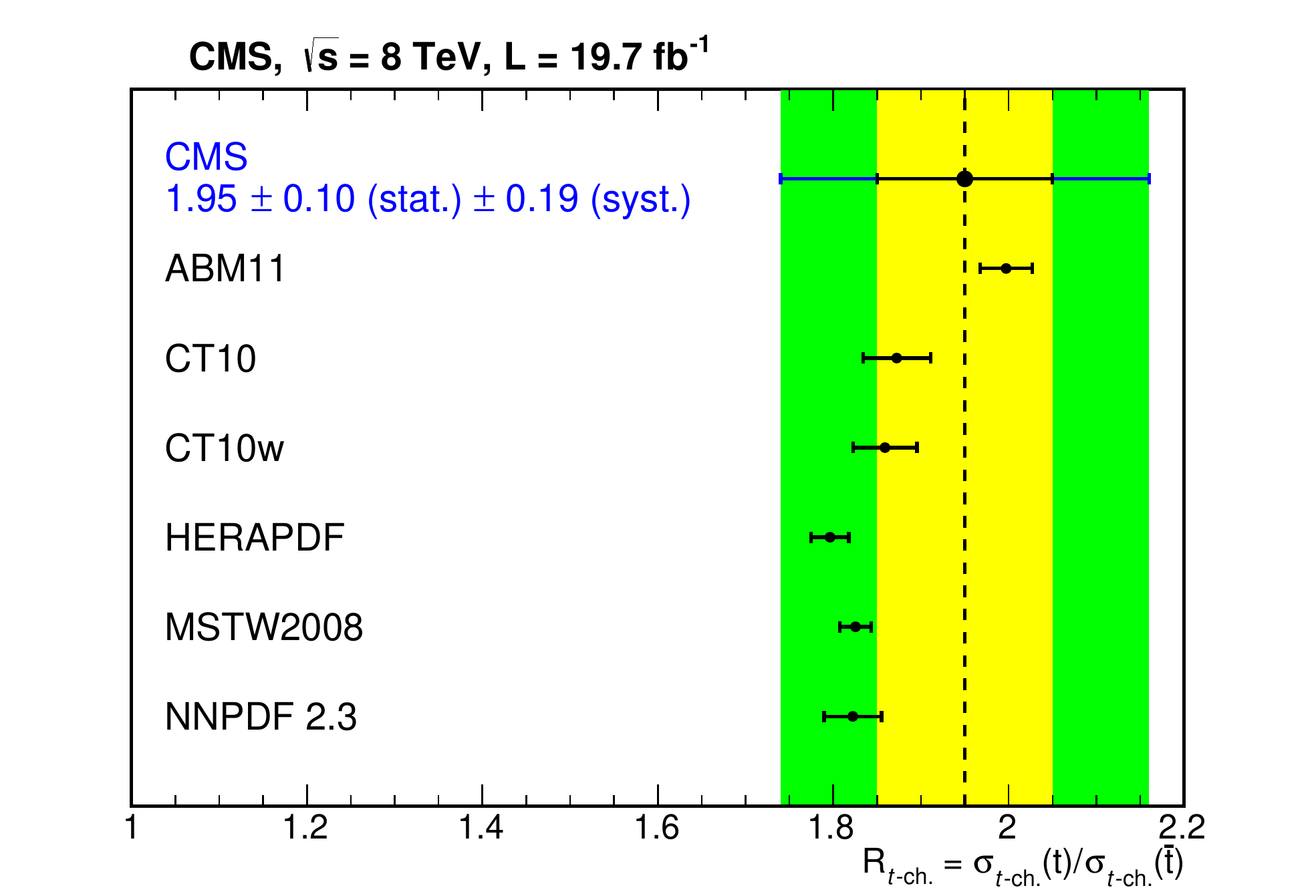}
\caption{2j1t distribution of the pseudorapidity of the untagged jet (top); $R$ value predictions for different PDFs (bottom), which are all in agreement with the measured value within the uncertainties. Taken from the CMS cross section measurement at $\sqrt{s}=8$\,TeV~\cite{cms8tev}.}
 \label{fig:8tevcms}
\end{figure}

The templates in the 2j1t region are simultaneously fit to data in $\mid\hspace{-0.085cm}\eta_\mathrm{j'}\hspace{-0.085cm}\mid$ in both the electron and muon channel. The (semi-)data driven background estimations explained above come with uncertainties that are reflected by nuisance parameters in the maximum-likelihood fit; the signal normalization is left free to float. The left distribution of Figure~\ref{fig:8tevcms} shows the post-fit distribution in the muon channel. A very good agreement between data and predictions is observed. The resulting cross section is $\sigma=83.6\pm2.3\,(\mathrm{stat.})\pm7.4\,(\mathrm{syst.})$\,pb. The dominating systematics are related to the modelling of the signal process and the jet energy scale. Separating the events by the lepton charge and fitting the top and anti-top templates independently gives $\sigma_\mathrm{t}=53.8\pm1.5\,(\mathrm{stat.})\pm4.4\,(\mathrm{syst.})$ and $\sigma_{\bar{\mathrm{t}}}=27.6\pm1.3\,(\mathrm{stat.})\pm3.7\,(\mathrm{syst.})$. Their ratio is $R=\sigma_\mathrm{t}/\sigma_{\bar{\mathrm{t}}}=1.95\pm0.10\,(\mathrm{stat.})\pm0.19\,(\mathrm{syst.})$. This pseudo-observable is sensitive to which PDF has been used in the calculation of the hard interaction. The right figure of Figure~\ref{fig:8tevcms} contrasts the measurement with different PDFs. The data does not really disfavor one of the sets within the uncertainties, but most of them predict a smaller $R$ value than the one observed. 

\subsection{Fiducial}

\begin{figure}
\includegraphics[width=0.38\textwidth]{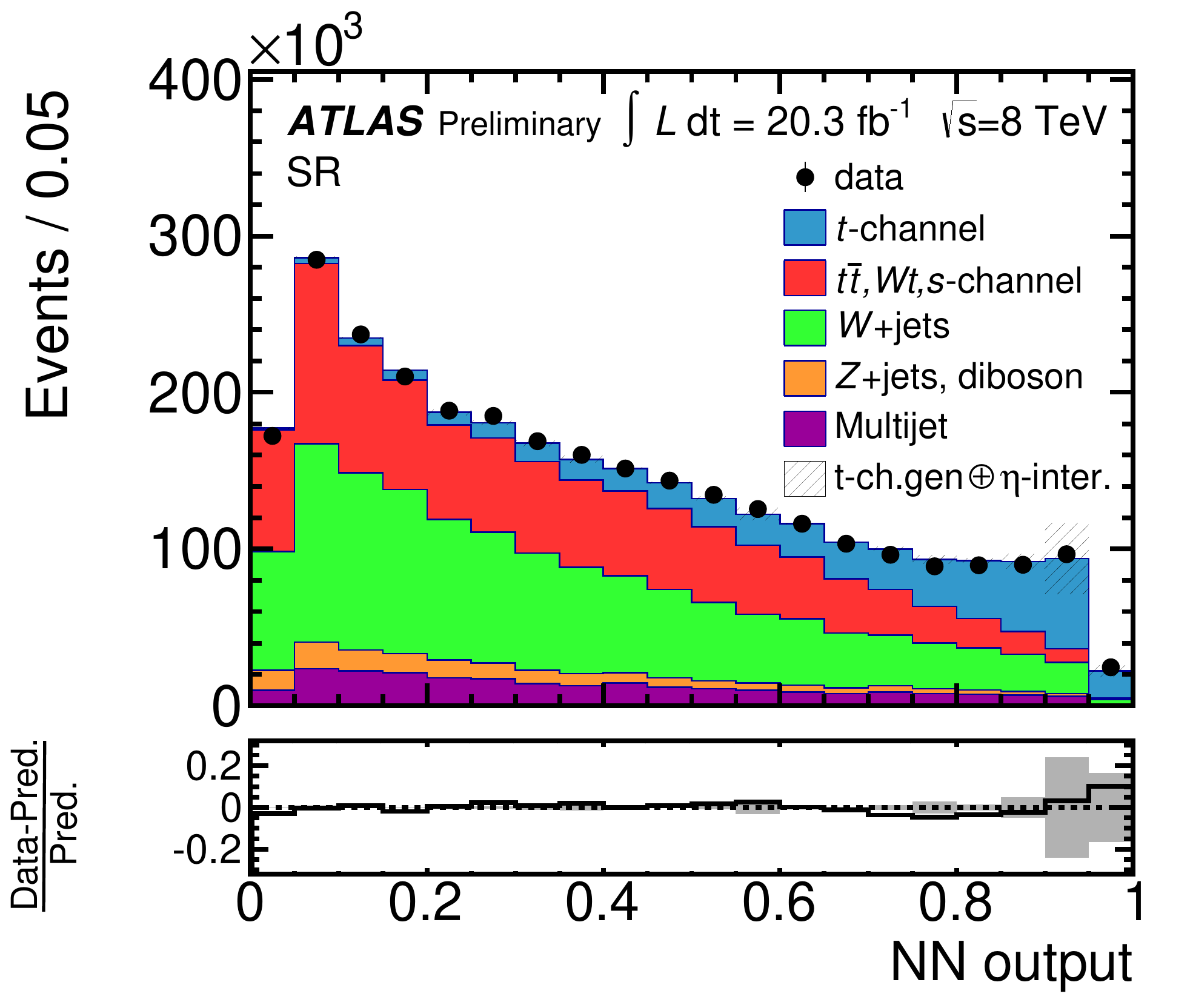}\hspace{1cm}
\includegraphics[width=0.38\textwidth]{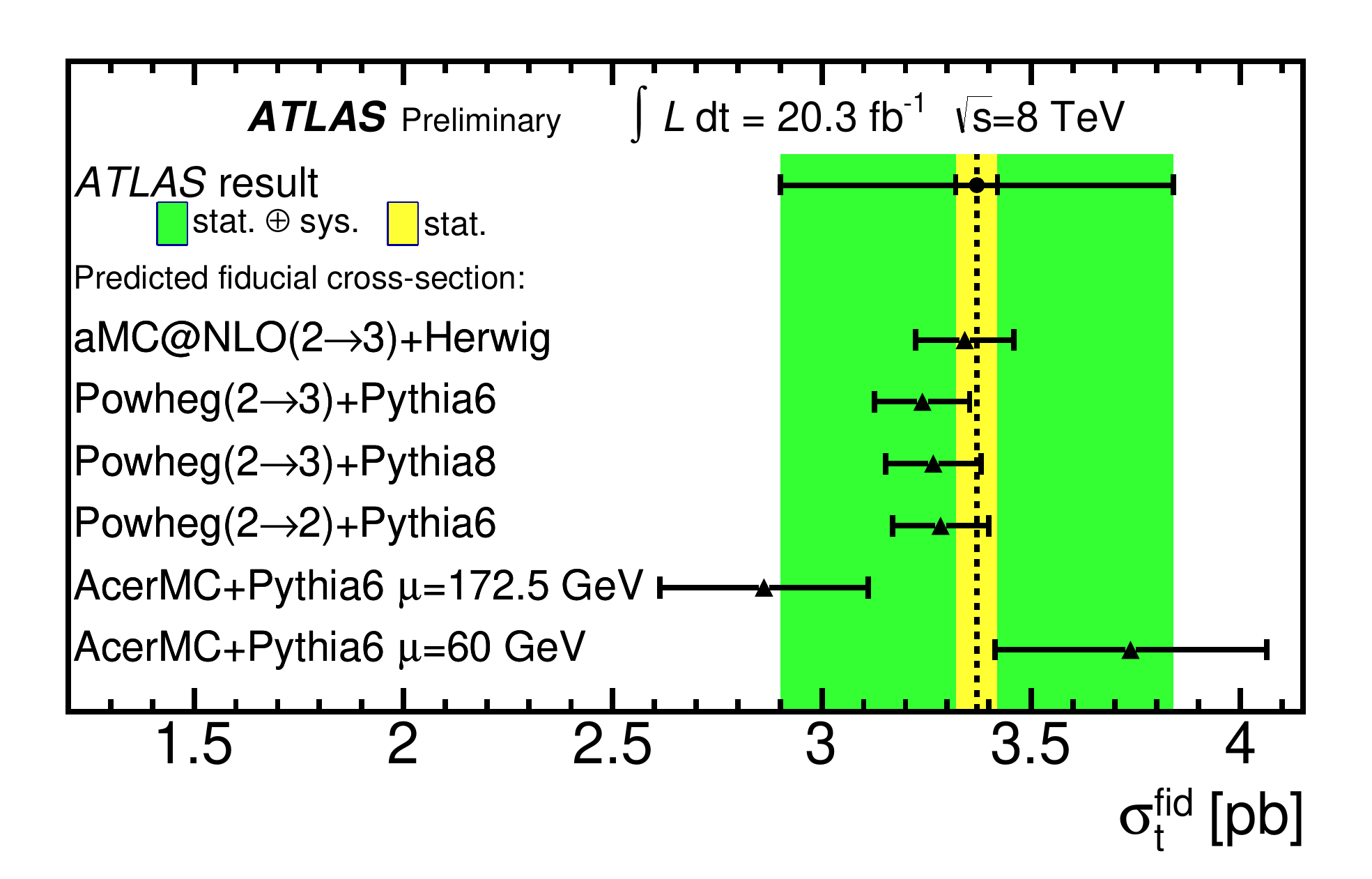}
\caption{Results from the ATLAS fiducial cross section measurement at $\sqrt{s}=8$\,TeV~\cite{atlas8tev}: post-fit NN output distribution in the 2j1t region (top). Fiducial cuts have been imposed on the simulations. The bottom figure compares the measured fiducial cross section with the predictions of various generation setups, disfavoring only the matched $2\to2$/$2\to3$ calculations.}
 \label{fig:8tevatlas}
\end{figure}

Compared to fully inclusive results, fiducial cross sections have the advantage that their dependence on the event generation (knobs to turn are e.g. matrix element generators, scale choices, hadronization models and PDFs) is reduced. Therefore differences related to the modelling are reduced to residual differences within the fiducial volume entirely covered by the experimental acceptance. Uncertainties stemming from the extrapolation from a visible into the inclusive phase space do not apply to such a measurement; moreover it is easier to re-interpret the results once better MC generators are available at some later point. Practically this is achieved my mimicking the selection imposed on reconstructed objects on behalf of cuts on stable particles at generator level. Consequently the following cuts are applied, defining a 2j1t signal region: exactly one lepton (muon or electron) with $\pt>25$\,GeV and $\abseta<2.5$. Jets are reconstructed within $\abseta<4.5$ and must have $\pt>30$\,GeV (or even $\pt>35$\,GeV if $2.75<\abseta<3.5$). Exactly two jets need to be present, one of which must be identified as a b~jet, either by deploying a multivariate algorithm for identifying secondary vertices on reconstruction level or by matching stable B hadrons to generated jets. The lepton must have a distance in the $\phi$-$\eta$-plane of $\Delta R=\sqrt{\Delta \phi^2+\Delta \eta^2}>0.4$ to any jet. QCD multi-jet events are rejected by requiring $E_\mathrm{T}^\mathrm{miss}>30$\,GeV  and $m_\mathrm{T}>50$\,GeV.

A Neural Network is trained with 14 variables, the three most relevant being the pseudorapidity of the untagged jet, the reconstructed top quark mass and the invariant mass of the jet pair. The shape of the discriminator is validated in a \ttbar~enriched 2j2t region and a 2j1t region with relaxed b~tagging requirement, which makes it being dominated by W + jets events. Except for the QCD multi-jet production, which is estimated in a data driven technique, shapes of all backgrounds are taken from simulation, and the templates are normalized to the most precise available (N)NLO theory predictions. In Figure~\ref{fig:8tevatlas} a good agreement between data and simulation is found after a maximum-likelihood fit has been performed in the Neural Network discriminator. The fit results translate into a measured fiducial cross section $\sigma_\mathrm{fid.}=3.37\pm0.05\,(\mathrm{stat.})\pm0.47\,(\mathrm{syst.})\pm0.09\,(\mathrm{lumi.})$. Figure~\ref{fig:8tevatlas} also shows a comparison of the result with predictions of various event generation setups, owing to the fact that, as mentioned in the introduction, the $t$-channel is a good place to constrain modelling aspects in the Monte-Carlo simulation. Except for the AcerMC setup, which is a calculation based on matching 4F and 5F events at leading order in $\alpha_\mathrm{S}$ based of the \pt~of the additional b quark, all setups give predictions that are well compatible with data. The inclusive cross section can easily be obtained -- at the cost of larger uncertainties due to the extrapolation -- by dividing the fiducial cross section by the selection effiency of the fiducial selection ($\sigma=(1/\epsilon_\mathrm{fid})\cdot\sigma_\mathrm{fid.}$) and turns out to be $\sigma=82.6\pm1.2\,(\mathrm{stat.})\pm11.4\,(\mathrm{syst.})\pm3.1\,(\mathrm{PDF})+2.3\,(\mathrm{lumi.})$. This information can be used to measure the CKM matrix element $V_\mathrm{tb}$, which is $\simeq1$ for the SM but whose value could be altered by new physics. Assuming $\mid\hspace{-0.085cm}V_\mathrm{tb}\hspace{-0.085cm}\mid\,\,\gg\mid\hspace{-0.085cm}V_\mathrm{ts}\hspace{-0.085cm}\mid,\mid\hspace{-0.085cm}V_\mathrm{td}\hspace{-0.085cm}\mid$ and $\mathcal{B}(\mathrm{t}\to\mathrm{bW})=1$, it is simply given by $\mid\hspace{-0.085cm}V_\mathrm{tb}\hspace{-0.085cm}\mid\,\,=\sqrt{\sigma/\sigma_\mathrm{theor.}}$ and numerically for this analysis $\mid\hspace{-0.085cm}V_\mathrm{tb}\hspace{-0.085cm}\mid\,\,=0.97^{+0.09}_{-0.10}\,(\mathrm{exp.+theor.})$, i.e. it is compatible with the SM prediction. More details are given in~\cite{atlas8tev}.

\subsection{Differential}

\begin{figure}
\includegraphics[width=0.34\textwidth]{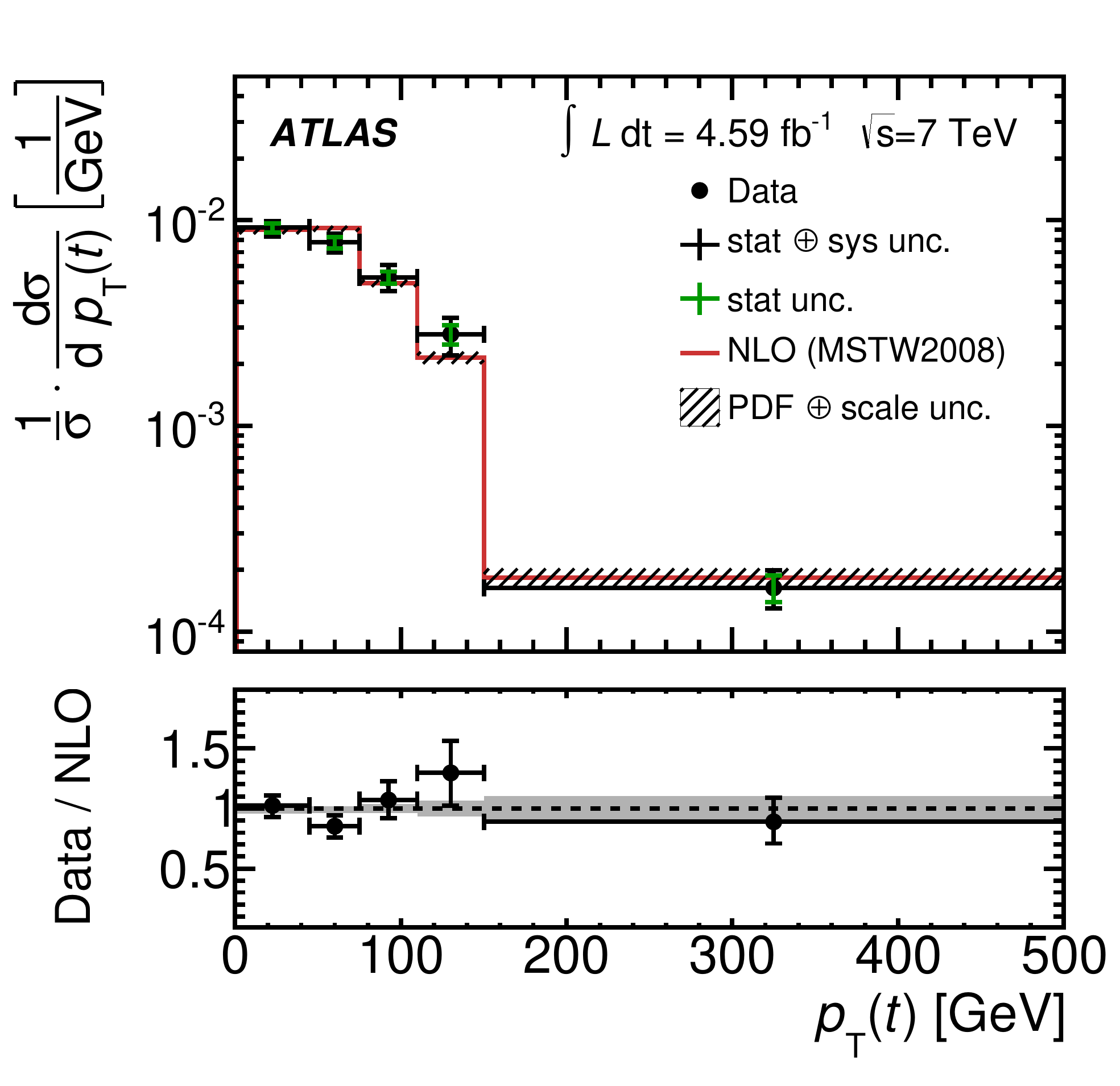}
\includegraphics[width=0.34\textwidth]{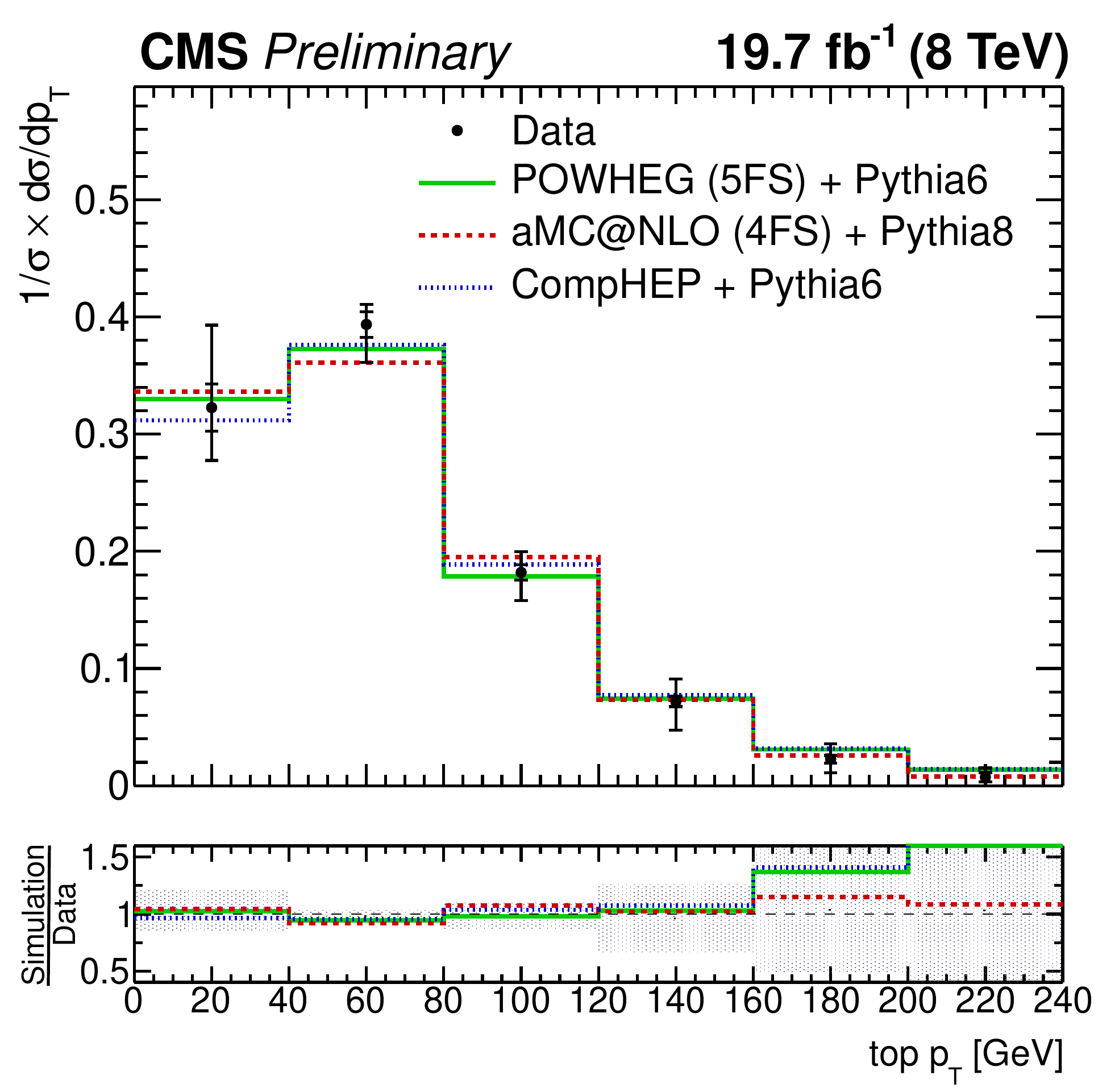}
\caption{Normalized distributions of the top quark \pt~have been measured by ATLAS (top) at $\sqrt{s}=7$ \cite{atlasdiff} and by CMS (bottom) at $8$\,TeV \cite{cmsdiff}.}
 \label{fig:unfolded}
\end{figure}

Both ATLAS and CMS have also measured a cross section differential in the $\pt$ of the top quark~\cite{atlasdiff,cmsdiff}. The analysis designs are similar: a Neural Network is trained  in the 2j1t region for a better separation between the $t$-channel and background processes, and its discriminator is cut on in order to obtain a high-purity sample of single top events. After the background contributions are subtracted, the distributions are unfolded to parton level, where the kinematics of the top quark are understood to reflect the resonance before its decay and after radiation effects. The normalized $\pt$ distributions are shown in Figure~\ref{fig:unfolded} for $\sqrt{s}=7$ and $8$\,TeV, respectively, and display good agreement between data and simulation. The (dis)agreement in the tail of the $8$\,TeV distribution suggests that the 4F is able to model high \pt~objects better than the 5F.

\section{W BOSON HELICITY}

The helicity of the W boson is usually subject to \ttbar~analyses and is measured with a single top selection for the first time in the analysis presented here \cite{helicity}. The helicity angle $\theta^*_\ell$ is defined as the angle between the direction of the reconstructed W boson in the top quark rest frame and the direction of the lepton in the W boson rest frame. Its probability function (which is the same for \ttbar~and single top events) is proportional to each helicity component ($F_\mathrm{L}$: left-handed, $F_0$: longitudinal, $F_\mathrm{R}$: right-handed) of the W boson,

\begin{align*}
&\frac{1}{\Gamma}\frac{\mathrm{d}\Gamma}{\mathrm{d}\cos{\theta^*_\ell}}=\frac{3}{8}(1-\cos{\theta^*_\ell})^2F_\mathrm{L}+\\
&\qquad\quad\qquad\frac{3}{4}\sin^2{\theta^*_\ell}F_0+\frac{3}{8}(1+\cos{\theta^*_\ell})^2F_\mathrm{R},
\end{align*}

where $\Gamma$ is the total width of the top quark decay. SM predictions are $F_\mathrm{L}=0.311\pm0.005$, $F_0=0.687\pm0.005$ and $F_\mathrm{R}=0.0017\pm0.0001$~\cite{czarnecki}. In the analysis they are extracted from a maximum-likelihood fit in $\cos{\theta^*_\ell}$.

\begin{figure}
\includegraphics[width=0.37\textwidth]{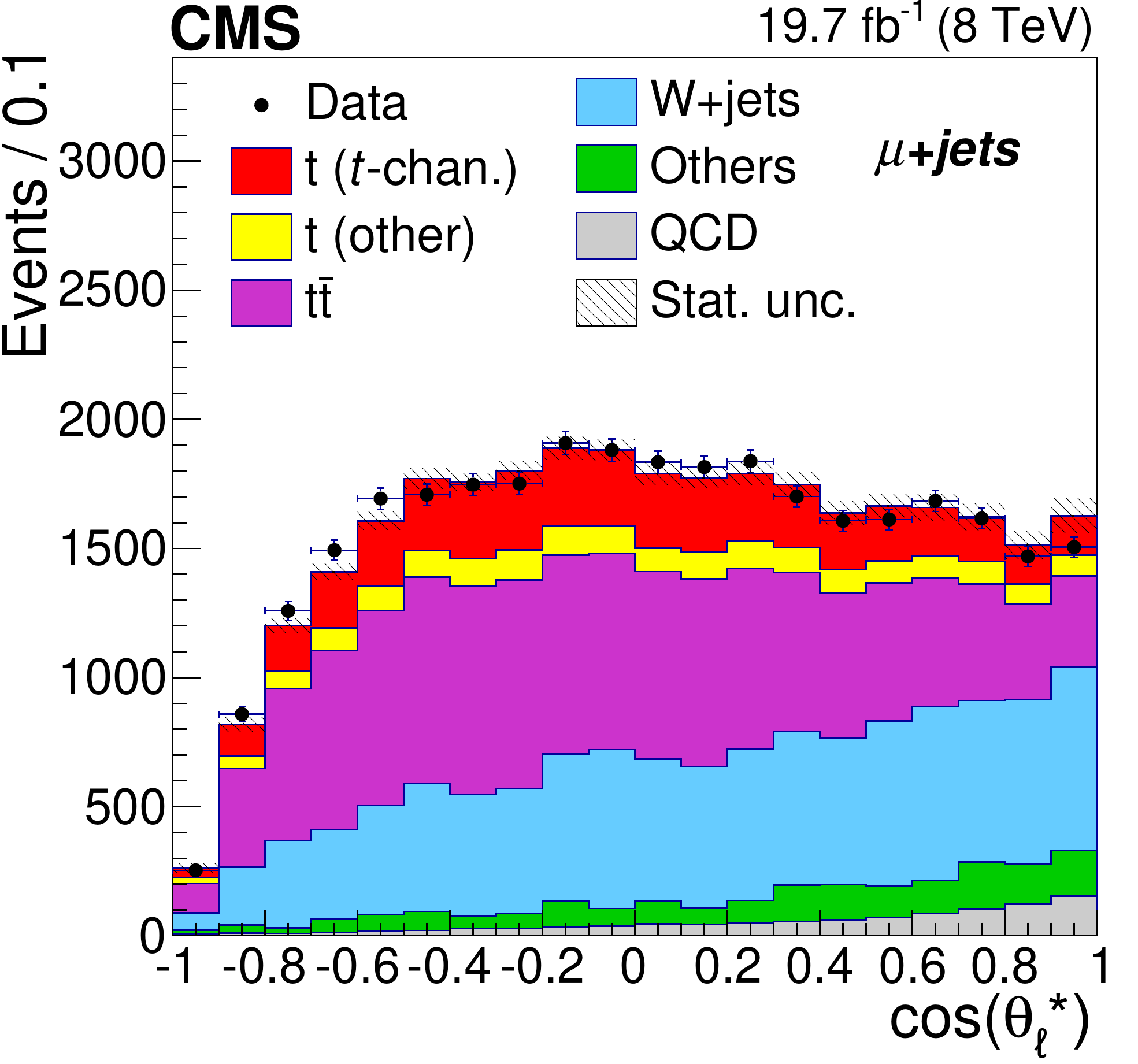}
\includegraphics[width=0.34\textwidth]{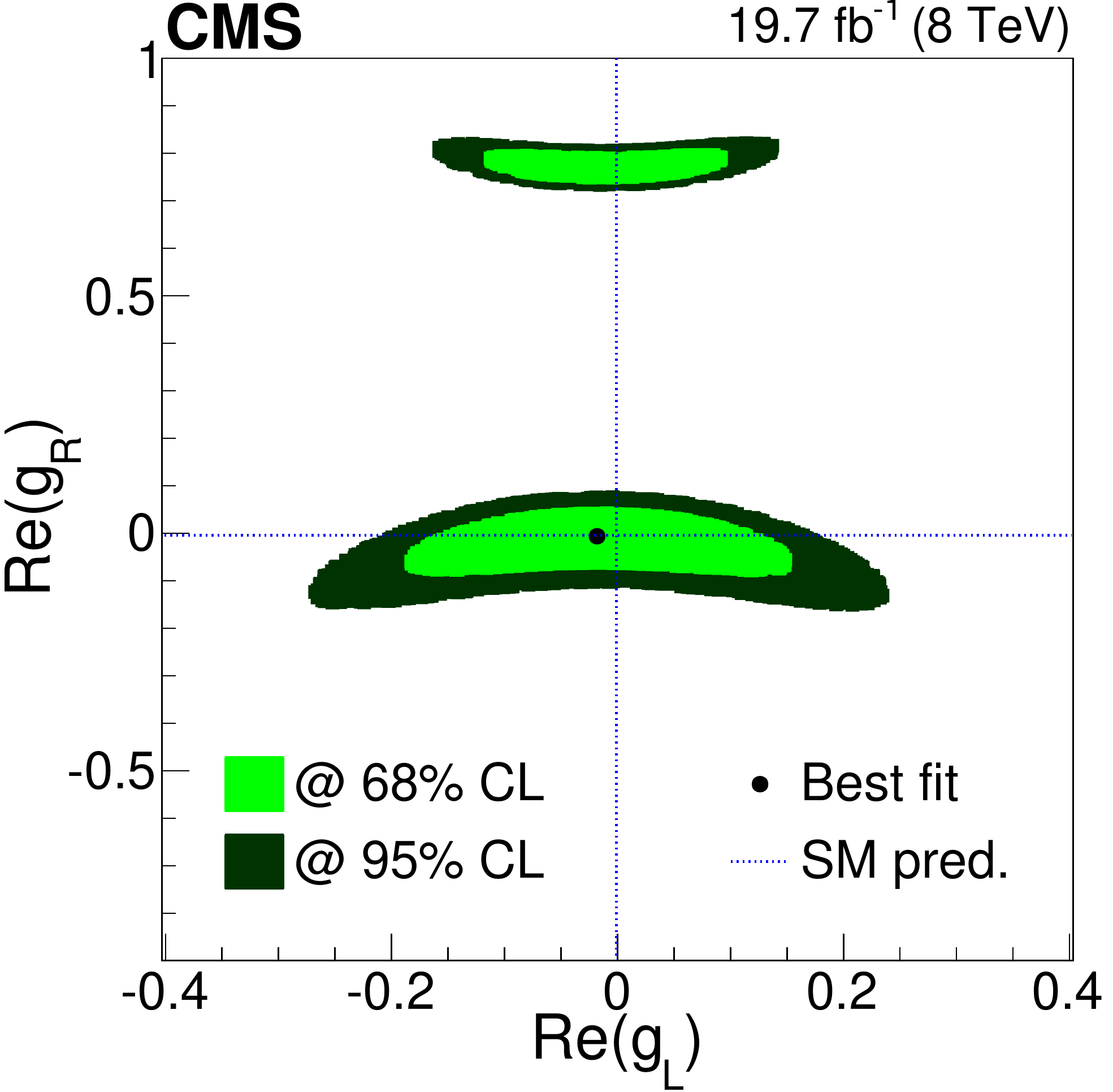}
\caption{Distribution of $\cos{\theta^*_\ell}$ (top) and limits on real tensor components in the tWb coupling (bottom), taken from the 8\,TeV CMS analysis~\cite{helicity}.}
 \label{fig:helicity}
\end{figure}

In terms of the event selection, it is closely following what has been done for the $t$-channel inclusive cross section measurement presented earlier. Since a boost in the top quark rest frame is required, a top quark candidate must be reconstructed. Two solutions for the $z$-component of the escaping neutrino arise when solving a quadratic equation for $p_{z,\upnu}$. Events which only have two imaginary solutions are discarded, otherwise the one with the smallest absolute value is picked. Figure~\ref{fig:helicity} shows the simulated $\cos{\theta^*_\ell}$ templates compared to data in the 2j1t muon channel. 
All single top events ($t$, $s$, tW) and \ttbar~events contribute to the signal sample, since one can reconstruct a tWb vertices in all of them. The shape in $\cos{\theta^*_\ell}$ of the main background, W + jets, is taken from simulation, while the normalization is introduced as an unconstrained parameter and is fit simultaneously together with two of the three helicity fractions, $F_\mathrm{L}$ and $F_0$. The third component is obtained from the constraint $\Sigma_i F_i=1$. Normalization estimates from \cite{cms8tev} are used for all other backgrounds, whose shapes are either taken from simulation or taken from a control region in the case of multi-jet production. The fit results are $F_\mathrm{L}=0.298\pm0.028\,(\mathrm{stat.})\pm0.032\,(\mathrm{syst.})$, $F_0=0.720\pm0.039\,(\mathrm{stat.})\pm0.037\,(\mathrm{syst.})$ and $F_\mathrm{R}=-0.018\pm0.019\,(\mathrm{stat.})\pm0.011\,(\mathrm{syst.})$, which is consistent with the SM predictions.

 The above results can be re-interpreted in order to exclude potential tensor terms appearing in the tWb couplings, whose real parts are given by the parameters $g_\mathrm{L}$ and $g_\mathrm{R}$ in the extended Lagrangian

\begin{align*}
&\mathcal{L}_\mathrm{tWb}^{\mathrm{anom.}}=-\frac{g}{\sqrt{2}}\bar{\mathrm{b}}\gamma^\mu(V_\mathrm{L}P_\mathrm{L}+V_\mathrm{R}P_\mathrm{R})\mathrm{tW}_\mu^- -\\
&\qquad\quad\qquad\frac{g}{\sqrt{2}}\bar{\mathrm{b}}\frac{i\sigma^{\mu\nu}{q}_\nu}{m_\mathrm{W}}(g_\mathrm{L}P_\mathrm{L}+g_\mathrm{R}P_\mathrm{R})\mathrm{tW}_\mu^- +\mathrm{h.c.},
\end{align*}

assuming a purely left-handed interaction of the vector part, i.e. $V_\mathrm{L}=1$, $V_\mathrm{R}=0$. The reader is deferred to the publication for further information and more details. The best fit values are $g_\mathrm{L}=-0.017$ and $g_\mathrm{R}=-0.008$. As can be seen in Figure~\ref{fig:helicity}, this is consistent with the leading order SM prediction of 0. The signal modelling is the main source of systematic uncertainty.

\section{TOP POLARIZATION}

\begin{figure}
\includegraphics[width=0.40\textwidth]{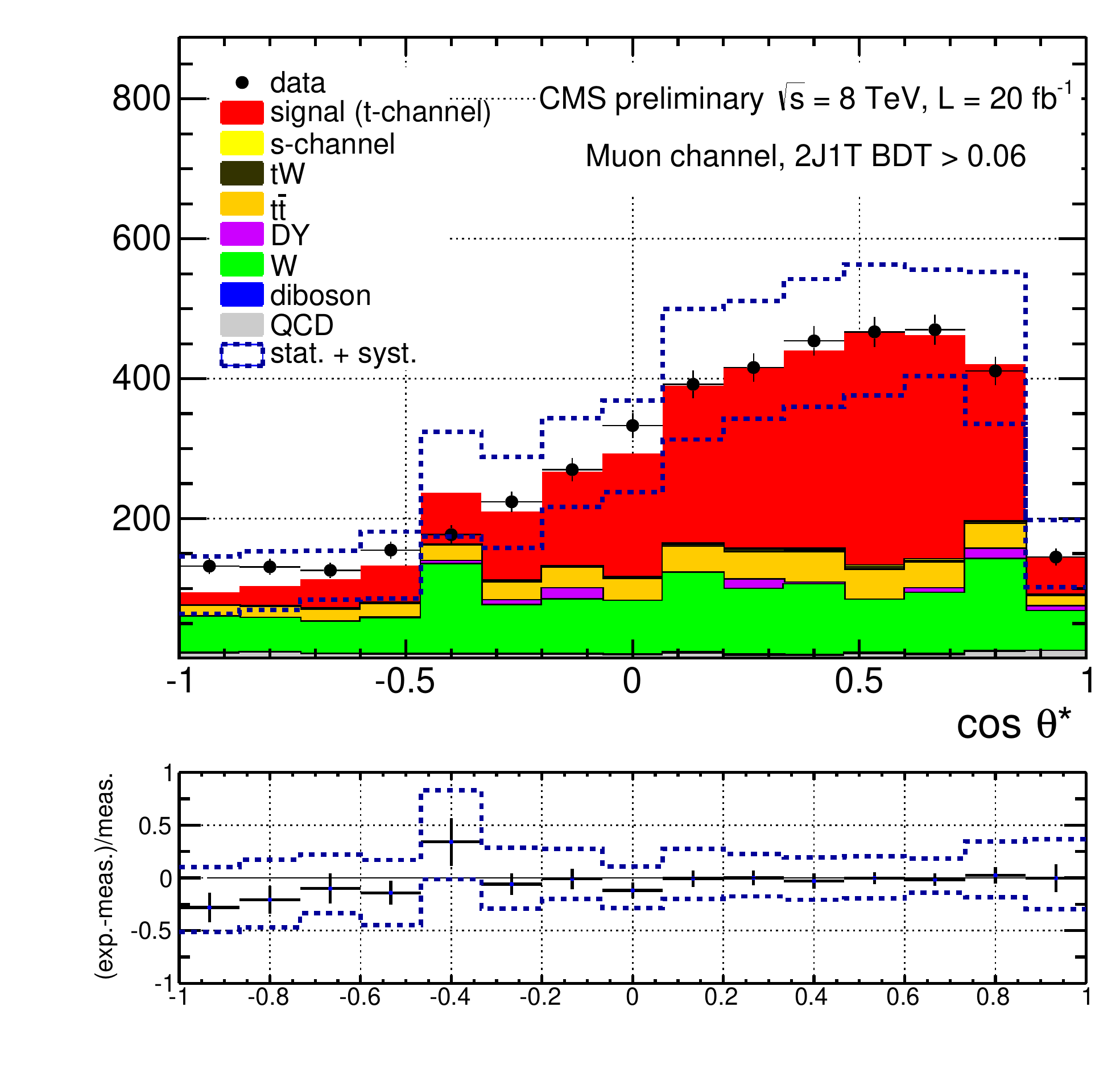}
\caption{The CMS analyis~\cite{polarization} measures the $\cos{\theta^*}$ distribution at $\sqrt{s}=8$\,TeV in the muon channel after a cut on the BDT output.}
 \label{fig:polarization}
\end{figure}

In the Standard Model the top quark is highly polarized along the direction of the light recoil quark, and its decay products bear information on the spin of the resonance. This can be used to construct an observable, $\cos{\theta^*}$, which is sensitive to the top quark polarization. It is defined as the angle between the lepton of the top quark decay and the light recoil jet, as seen from the top quark rest frame. Its differential distribution follows

\begin{equation}
\frac{1}{\Gamma}\frac{\mathrm{d}\Gamma}{\mathrm{d}\cos{\theta^*}}=\frac{1}{2}(1+P_\mathrm{t}\alpha_\ell\cos{\theta^*}),
\end{equation}

where $P_\ell$ denotes the top quark polarization is $\alpha_\ell$ is the degree of correlation of the angular distribution of $\ell$ with respect to the top quark spin (this analysis assumes $\alpha_\ell=1$). The measured quantity finally is the asymmetry $A_\ell=(N(\cos{\theta^*}>0)-N(\cos{\theta^*}<0))/(N(\cos{\theta^*}>0)+N(\cos{\theta^*}<0))$, which is determined separately in both the electron and muon channel. A more detailed description of the applied cuts that enhance signal over background contributions is provided in~\cite{polarization}, but it employs a 2j1t selection similar to~\cite{cms8tev}. The multi-jet background is derived from data in a control region enriched in QCD events that is obtained by inverting lepton isolation criteria. The default simulation setup for W + jets suffers from a bad description of $\cos{\theta^* }$ close to $-1$; another generator (Sherpa) which performs better in this region of phase space is used to correct the main MadGraph simulation. The shapes of \ttbar~templates are validated in a 3j1t and 3j2t control region. Finally a boosted decision tree (BDT) is trained in order to further separate single top production from the backgrounds. After a maximum-likelihood fit to its shape, in which the signal and background normalizations are determined, a cut on the BDT output is imposed to obtain a signal-enriched sample (see Figure~\ref{fig:polarization}). The background contributions are subtracted from data, and the distribution is unfolded to correct for detector effects. The asymmetry $A_\ell$ is calculated from the unfolded templates separately for the electron and the muon channel. Their combination gives $A_\ell=0.41\pm0.06\,(\mathrm{stat.})\pm0.07\,(\mathrm{syst.})$, where the jet energy scale is the dominant source of systematic uncertainty. This results in a top quark polarization of $P_\mathrm{t}=0.82\pm0.12\,(\mathrm{stat.})\pm0.32\,(\mathrm{syst.})$.

\end{document}